\documentclass{aa}
\usepackage{graphicx}
\usepackage[varg]{txfonts}
\usepackage{cases}
\usepackage{natbib}

\begin{document}

\title{Long-period energy releases during a C2.8 flare}

\author{Dong~Li\inst{1,2}, Jianping~Li\inst{1}, Jinhua~Shen\inst{3}, Qiwu~Song\inst{1}, Haisheng~Ji\inst{1}, and Zongjun~Ning\inst{1}}

\institute{Key Laboratory of Dark Matter and Space Astronomy, Purple Mountain Observatory, Chinese Academy of Sciences, Nanjing 210023, PR China \email{lidong@pmo.ac.cn} \\
           \and State Key Laboratory of Space Weather, Chinese Academy of Sciences, Beijing 100190, PR China \\
           \and Xinjiang Astronomical Observatory, Chinese Academy of Sciences, Urumqi, 830011, PR China \\
           }
\date{Received; accepted}

\titlerunning{Long-period energy releases during a C2.8 flare}
\authorrunning{Li et al.}

\abstract {The study of quasi-periodic pulsations (QPPs) is a key
diagnostic of intermittent or periodic energy releases during solar
flares.} {We investigated the intermittent energy-releasing
processes by analyzing the long-period pulsations during a C2.8
flare on 2023 June 03.} {The solar flare was simultaneously observed
by the solar X-ray detector  on board the Macau Science
Satellite-1B, the Geostationary Operational Environmental Satellite,
the Chinese H$\alpha$ Solar Explorer, the Expanded Owens Valley
Solar Array, the Atmospheric Imaging Assembly, and the Extreme
Ultraviolet Variability Experiment for the Solar Dynamics
Observatory.}{The C2.8 flare shows three successive and repetitive
pulsations in soft X-ray (SXR) and high-temperature extreme
ultraviolet (EUV) emissions, which may imply three episodes of
energy releases during the solar flare. The QPP period is estimated
to be as long as $\sim$7.5~minutes. EUV imaging observations suggest
that these three pulsations come from the same flare area dominated
by the hot loop system. Conversely, the flare radiation in
wavelengths of radio/microwave, low-temperature EUV, ultraviolet
(UV), and H$\alpha$ only reveals the first pulsation, which may be
associated with nonthermal electrons accelerated by magnetic
reconnection. The other two pulsations in wavelengths of SXR and
high-temperature EUV might be caused by the loop-loop interaction.}
{Our observations indicate that the three episodes of energy
releases during the C2.8 flare are triggered by different
mechanisms, namely the accelerated electron via magnetic
reconnection, and the loop-loop interaction in a complicated
magnetic configuration.}


\maketitle

\section{Introduction}
A solar flare is an explosive and impulsive energy-releasing
phenomenon via the well-known magnetic reconnection, which can be
detected in multi-height solar atmospheres \citep{Benz17}. The
released energy is mainly converted from the free magnetic energy
that is stored in the nonpotential magnetic field, which overlaps
the current-carrying field and is dominated by a complex magnetic
geometry \citep{Jiang21,Yan22,McKevitt24}. As described in the
classical two-dimensional (2D) flare model \citep{Priest02}, the
free magnetic energy may release violently and suddenly in the solar
coronal active region, which can rapidly heat thermal plasmas to the
 temperature of several million Kelvin (MK), and they can quickly
accelerate nonthermal particles to a high energy range of dozens of
keV to GeV \citep[e.g.,][]{Warmuth16,Li21a}. In the imaging
observation for a typical solar flare, double footpoints are seen in
hard X-ray (HXR) or microwave emissions, and a loop-like structure
that connects double footpoints appears in soft X-ray (SXR) and
extreme ultraviolet (EUV) wavelengths, while the loop-top source
might be observed in HXR or microwave channels
\citep[e.g.,][]{Masuda94,Yan18,Li23}. Meanwhile, two ribbon-like
features can be found in wavelengths of ultraviolet (UV), white
light (WL), or H$\alpha$ \citep{Li17,Tian17}. This is  called a
two-ribbon flare, which is the most common flare and matches the
standard 2D reconnection model \citep{Priest02}. If the magnetic
configuration becomes more complicated, then the three-ribbon flare
\citep[e.g.,][]{Wang14,Zimovets21a} or circular-ribbon flare
\cite[e.g.,][]{Masson09,Ning22} are observed on the Sun, which
may be associated with various 3D magnetic configurations
\citep{Janvier15,Zhang24}.

A solar flare is commonly powered by the release process of free
magnetic energy \citep{Priest02,Shen23}. Such an energy release is
often characterized by repeated and successive processes, which is
manifested as quasi-periodic pulsations (QPPs) in the time series of
flare radiation \citep[][and references
therein]{Nakariakov19,Zimovets21b}. The typical QPP often has at
least three successive and complete peaks, namely three QPP cycles.
If there are only one or two peaks, it might   just be a
coincidence; that is, the similar time interval between successive
peaks occurred by chance \citep[e.g.,][]{McLaughlin18,Li22f}. A QPP
event usually carries the abundant information of time
characteristics and plasma features in the flare source, and thus it
plays a crucial role in diagnosing the coronal parameter of the Sun
or remote solar-like stars \citep{Yuan19,Kolotkov21,Li21,Inglis23}.
The flare QPPs were first reported by \cite{Parks69} in
temporal-intensity profiles at X-ray and radio emissions. To date,
they have been observed in a broad wavelength range of
electromagnetic radiation: radio/microwave, H$\alpha$, Ly$\alpha$,
UV/EUV, SXR/HXR, and gamma-rays
\citep[e.g.,][]{Brosius15,Nakariakov18,Chen19,Hayes19,Li20a,Li20b,Milligan20,Hong21,Lid22,Motyk23}.
Their periods were measured from subseconds through seconds to
several minutes
\citep[e.g.,][]{Tian08,Samanta21,Shen22,Zimovets22,Li23b,Collier23,Mehta23,Zhao23,Zhou24}.
It seems that the detected periods were always dependent on the
studied wavebands or flare phases, indicating that flare QPPs
belonging to different categories may be generated by various
mechanisms or models \citep{Kupriyanova20}. There was not a single
mechanism that can fully interpret the presence of all flare QPPs
because there was not enough observational data  to distinguish
between the various mechanisms \citep{Inglis23}. The flare QPPs at
periods of
 subseconds were always measured in radio and microwave emissions,
which were usually produced by dynamic interactions between
energetic ions and electromagnetic waves that were trapped in complex
magnetic fields \citep{Aschwanden87,Tan12}. The flare QPPs at long
periods of seconds or minutes can be observed in nearly   all the
wavebands such as radio, H$\alpha$, L$\alpha$, UV/EUV, SXR/HXR, and
they are commonly associated with the magnetohydrodynamic (MHD)
waves in various modes \cite[see][]{Nakariakov20}.
Moreover, the flare QPPs detected in wavelengths of HXR and microwave
emissions during the impulsive phase were frequently interpreted in
terms of nonthermal electrons periodically accelerated by repetitive
magnetic reconnections
\citep{Farhang22,Li22,Karampelas23,Karlicky23,Corchado24}.

One key signature of the flare energy-releasing process is
impulsive, repetitive, or intermittent, which may be strongly
dependent on the time profiles on timescales of minutes, seconds,
and milliseconds \cite[cf.][]{Inglis23}. Such signatures of energy
release are often associated with   the flare radiation observed in
multiple wavelengths that is bursty and periodic; in other words,
the flare radiation is frequently modulated by a pattern of QPPs.
Therefore, the study of flare QPPs is a crucial diagnostic of the
intermittent energy releases on the Sun. For this article we
investigated the intermittent energy releases with   long periods
during the C2.8 flare on 2023 June 03. The article is organized as
follows: Section~2 introduces the observations and instruments,
Section~3 shows the data analysis and our main results, and
Section~4 presents some discussion, while a brief summary is given
in Section~5.

\section{Observations and Instruments}
The targeted solar flare occurred in the active region of
NOAA~13319. It was a C2.8 class flare, that started at
$\sim$21:40~UT on 2023 June 03, peaked at $\sim$21:47~UT, and
stopped at about
$\sim$22:10~UT.\footnote{https://www.solarmonitor.org/?date=20230603}
The C2.8 flare was simultaneously measured by multiple space- and
ground-based instruments: the solar X-ray detector (SXD) on board
the Macau Science Satellite-1B \citep[MSS-1B;][]{Shi23}; the
Geostationary Operational Environmental Satellite
\citep[GOES;][]{Chamberlin09}; the Large-Yield RAdiometer
\citep[LYRA;][]{Dominique13} on board the PRoject for On-Board
Autonomy~2 (PROBA2); the Atmospheric Imaging Assembly
\citep[AIA;][]{Lemen12} and the EUV SpectroPhotometer
\citep[ESP;][]{Didkovsky12} for the Solar Dynamics Observatory
(SDO); the Chinese H$\alpha$ Solar Explorer \citep[CHASE;][]{Lic22};
the STEREO/WAVES instrument \citep[SWAVES;][]{Kaiser08}; the
Expanded Owens Valley Solar Array \citep[EOVSA;][]{Gary11}; and the
e-CALLISTO radio spectrograph at NORWAY.

The Macau Science Satellite-1 (MSS-1) includes two subsatellites,
named   MSS-1A and MSS-1B. SXD on board MSS-1B is designed to
monitor the full-disk solar spectrum in the SXR/HXR energy range of
about 1$-$600~keV (or $\sim$0.02$-$12~{\AA}), and thus it has two
parts: the SXR detection unit (SXDU) and the HXR detection unit
(HXDU). The time cadence is as high as $\sim$1~s. GOES records the
full-disk solar irradiance in SXR 1$-$8~{\AA} and 0.5$-$4~{\AA} with
a time cadence of 1~s, and the isothermal temperature of
SXR-emitting plasma can be determined from their ratio. SDO/ESP
provides the full-disk solar irradiance with a time cadence of
0.25~s, which contains one SXR and four EUV wavebands. LYRA measures
the full-disk solar radiation in two far-ultraviolet (FUV) and two
X-ray ultraviolet (XUV) wavebands with a time cadence of 0.05~s.
SDO/AIA simultaneously measures the entire-Sun maps in seven EUV and
two UV wavelengths with a spatial scale of 0.6$^{\prime\prime}$ via
a standard pre-process. The time cadence for the EUV waveband is
12~s, and that for the UV wavelengths is 24~s. CHASE takes the
spectroscopic observation of the whole Sun in passbands of H$\alpha$
with a pixel scale of 1.04$^{\prime\prime}$, and a time cadence of
about 71~s. EOVSA is a radioheliograph, and it acquires the solar
dynamic spectrum in the microwave frequency of 1$-$18~GHz with a
time cadence of roughly 1~s. The e-CALLISTO radio spectrograph at
NORWAY observes the solar radio spectrum in the frequency range of
$\sim$1.0$-$1.6~GHz with a time cadence of $\sim$0.5~s. SWAVES
provides the solar radio spectrum at   frequencies of about
0.0026$-$16.025~MHz with a time cadence of 60~s.

\section{Data analysis and results}
\begin{figure}
\centering \noindent\includegraphics[width=\linewidth]{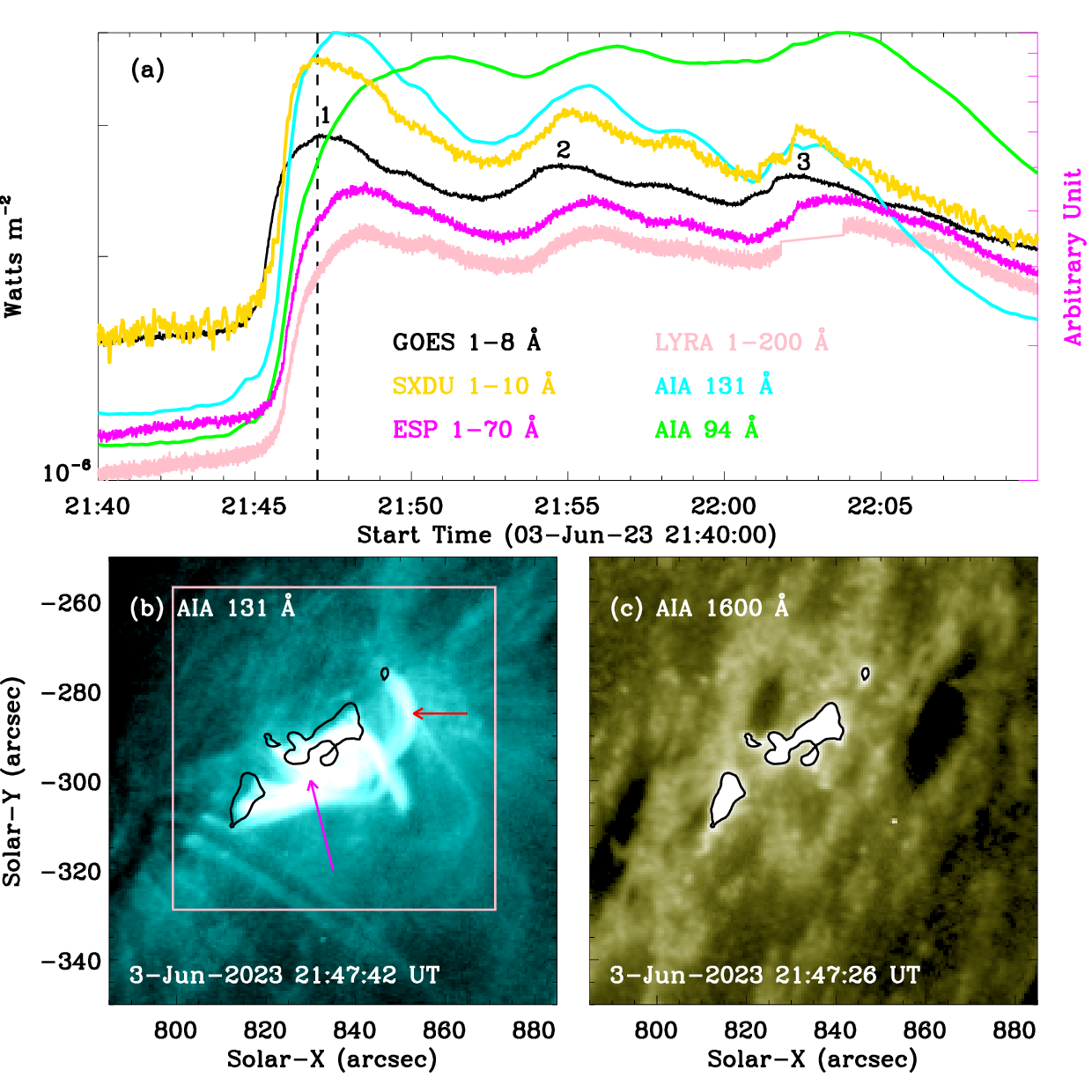}
\caption{Overview of the C2.8 flare on 2023 June 03. Panel (a):
Full-disk light curves recorded by GOES~1$-$8~{\AA} (black),
SXDU~1-10~{\AA} (gold), ESP~1$-$70~{\AA} (magenta), and
LYRA~1$-$200~{\AA} (pink), as well as the local fluxes measured by
SDO/AIA at 131~{\AA} (cyan) and 94~{\AA} (green). The vertical line
marks the flare peak time. Panels (b) and (c): EUV/UV snapshots with
a FOV of $\sim$100$^{\prime\prime}$$\times$100$^{\prime\prime}$ in
wavelengths of AIA~131~{\AA} and 1600~{\AA}. The black contours
outline flare ribbons in AIA~1600~{\AA}. The colored arrows indicate
hot loop systems. The pink box marks the flare region
($\sim$72$^{\prime\prime}$$\times$72$^{\prime\prime}$) used to
integrate the local flux. A movie associated to this figure is
available online. \label{over}}
\end{figure}

Figure~\ref{over} presents the  multiwavelength observations of the
targeted flare on 2023 June 03. The whole evolution of the flare can
be seen in the online animation. Figure~\ref{over}~(a) shows the
SXR/XUV light curves integrated over the whole Sun between 21:40~UT
and 22:10~UT. The GOES flux at 1$-$8~{\AA} (black) indicates a C2.8
class flare that reached its maximum at about 21:47~UT, as marked by
the dashed vertical line. The C2.8 flare also reveals another two
pulsations (labeled 2 and 3), which occurred at about 21:55~UT and
22:02~UT in the GOES~1$-$8~{\AA} flux, respectively, following
closely after the first pulsation (labeled 1). The three successive
pulsations might be regarded as a flare QPP with an average period
of about 7.5~minutes. These three successive pulsations can also be
seen in the SXR/XUV light curves recorded by SXDU~1$-$10~{\AA},
ESP~1$-$70~{\AA}, and LYRA~1$-$200~{\AA}, suggesting that the QPP
feature is visible by all the available instruments. We also note
that the longer the observed wavelengths, the later the peak times,
which means that the SXR flux at ESP~1$-$70~{\AA} (magenta) is
clearly later than that at SXDU~1$-$10~{\AA} (gold). This is because
the different peak times could have to do with different responses
of the instruments to the temperature of the plasma. The SXR/XUV
fluxes are integrated over the entire Sun, making it difficult to
determine whether these three pulsations come from the same flare
area. The SDO/AIA data can spatially resolve the flare region from
the whole Sun, and the local light curves in EUV wavebands are
integrated over the active region, as outlined by the pink box in
panel~(b). It can be seen that the local light curves at
AIA~131~{\AA} (cyan) and 94~{\AA} also show three successive
pulsations, confirming that they   originate from the same active
region. Panels~(b) and (c) show AIA submaps with a field of view
(FOV) of $\sim$100$^{\prime\prime}$$\times$100$^{\prime\prime}$. The
flare reveals some ribbons in the AIA~1600~{\AA} image, as outlined
by the black contours. These ribbons are connected by two groups of
hot loop systems, as can be seen in the AIA~131~{\AA} image.

\begin{figure}
\centering \noindent\includegraphics[width=\linewidth]{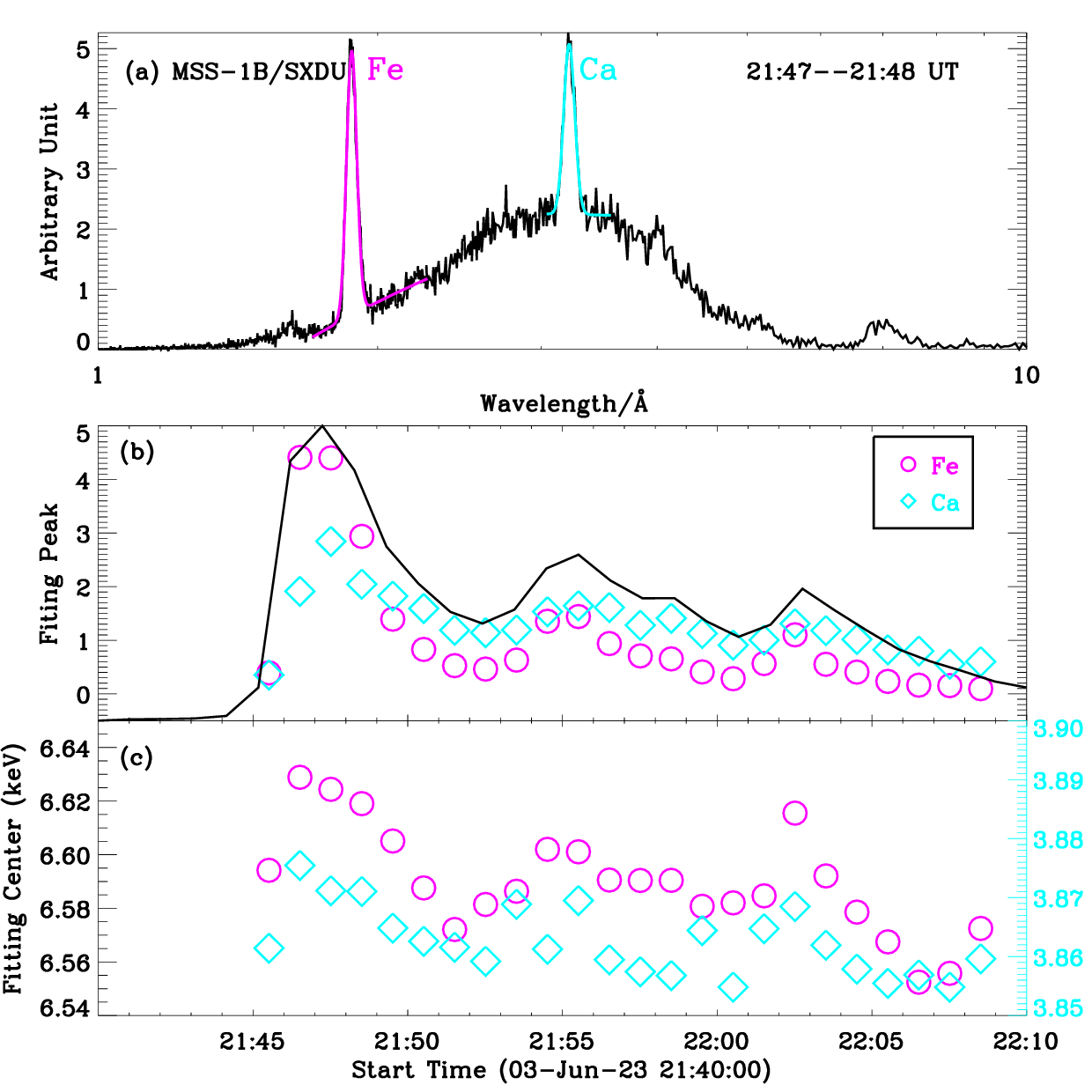}
\caption{Observational results measured from MSS-1B/SXDU. Panel (a):
SXR spectrum integrated over one~minute during the C2.8 flare in the
range 1$-$10~{\AA}. The magenta and cyan curves represent the
Gaussian fitting results for the Fe and Ca lines, respectively.
Panels (b) and (c): Time series of the fitting peak and center for
Fe (circular) and Ca (diamond) lines. \label{sxd}}
\end{figure}

The SXR light curves recorded by SXDU~1$-$10~{\AA} and
GOES~1$-$8~{\AA} appear to match  each other, confirming that the
newly launched MSS-1B/SXD can also be applied to monitor the solar
irradiance. Figure~\ref{sxd}~(a) shows the energy spectrum in the
SXR range of 1$-$10~{\AA} that was captured by MSS-1B/SXDU. In order
to improve the signal-to-noise ratio, the energy spectrum was
integrated over 60~s during the C2.8 flare (i.e., during
21:47$-$21:48~UT). It is immediately clear  that  the SXR energy
spectrum is mainly composed of two spectral lines superposed on a
strong background profile. The two spectral lines contains a group
of lines of highly stripped iron (Fe) and calcium (Ca) ions
\citep{Phillips04}. They both show Gaussian profiles, and a single
Gaussian function is applied to fit, as indicated by the magenta and
cyan curves, respectively. Then, their fitting peak intensities and
center positions are extracted, as shown in Figure~\ref{sxd}~(b) and
(c). The black line in panel~(b) represents the SXR flux recorded by
SXDU~1$-$10~{\AA} with a time cadence of 60~s. All these curves,
such as the fitting peak and center of Fe and Ca, reveal three
successive pulsations, similarly to what was observed in the SXR
flux recorded by GOES~1$-$8~{\AA}.

\begin{figure}
\centering \noindent\includegraphics[width=\linewidth]{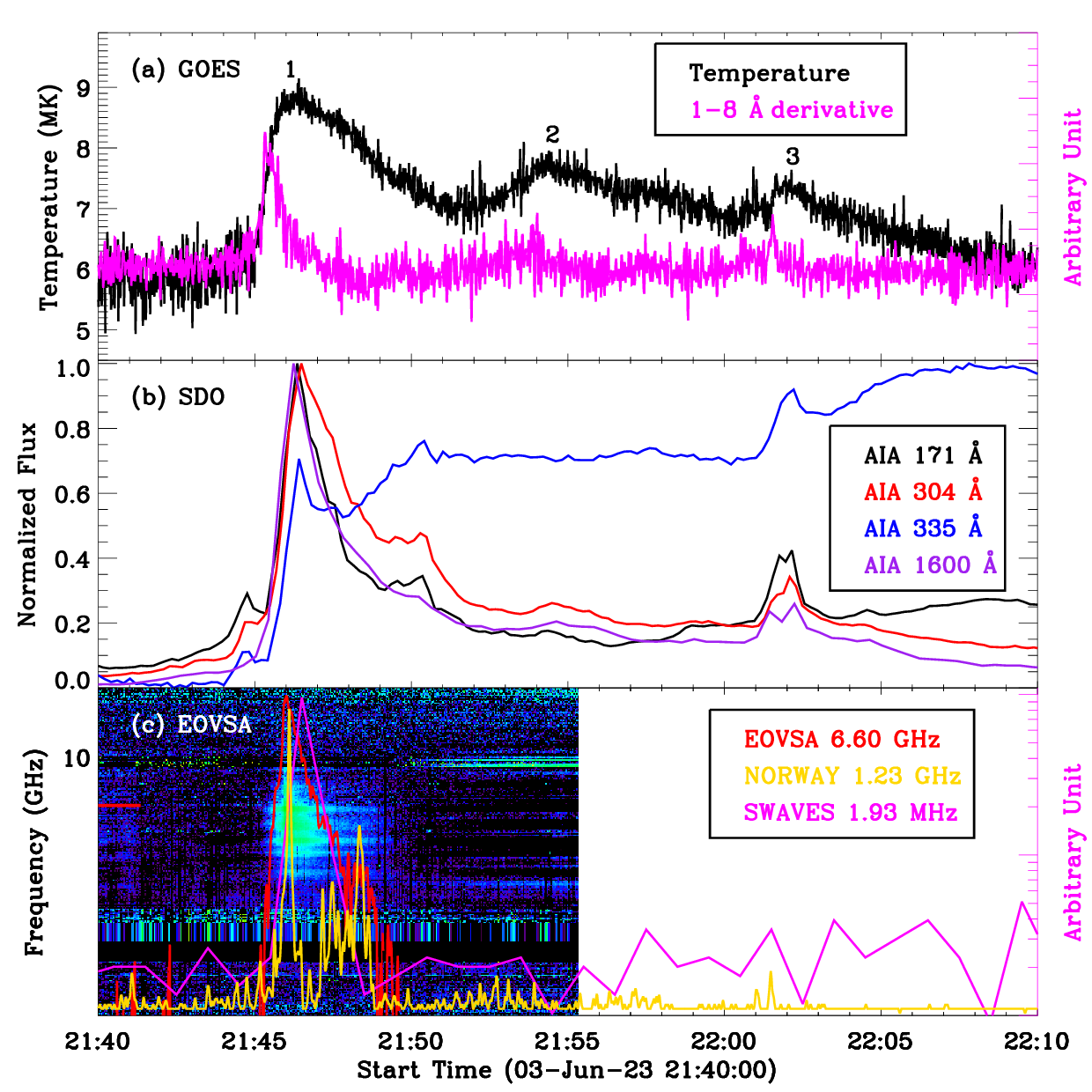}
\caption{Multiwavelength light curves during the C2.8 flare. Panel
(a): Full-disk light curves for the GOES temperature (black) and
SXR~1$-$8~{\AA} derivative (magenta). Panel (b): Local fluxes in
wavelengths of AIA~171~{\AA} (black), 304~{\AA} (red), 335~{\AA}
(blue), and 1600~{\AA} (purple). Panel (c) Radio light curves in
frequencies of EOVSA 6.60~GHz (red), NORWAY~1.23~GHz (gold), and
SWAVES~1.93~MHz (magenta). The context image is the radio dynamic
spectrum measured by EOVSA. \label{flux}}
\end{figure}

To investigate the origin or driver of the flare QPP that was
simultaneously detected in wavelengths of SXR, high-temperature EUV,
and spectral lines, Figure~\ref{flux} presents more light curves
during the C2.8 flare. Panel~(a) plots the time series of isothermal
temperature and SXR derivative that were derived from two GOES
fluxes. Similar to what was seen in the SXR fluxes, the temperature
profile also exhibits three main pulsations, suggesting that the
flare QPP is sensitive to the plasma temperature. Conversely, the
SXR 1$-$8 derivative flux only reveals clearly the first pulsation
(1), the other two pulsations (2 and 3) are very weak and even
invisible, which may indicate that the first pulsation is highly
associated with the nonthermal electron. Panel~(b) shows the local
EUV/UV fluxes (pink box in Figure~\ref{over}) in wavelengths of
AIA~171~{\AA}, 304~{\AA}, 335~{\AA}, and 1600~{\AA}. They all have
the first pulsation, but the other two pulsations are invisible (2)
or weak (3). The AIA~335~{\AA} flux appears to continually enhance
after the first pulsation, which is absolutely different from the
previous observation in SXR channels. Panel~(c) shows the radio
fluxes in frequencies of EOVSA~6.20~GHz, NORWAY~1.23~GHz, and
SWAVES~1.93~MHz. We can see that they all show one pulsation at
about 21:46~UT, which is consistent with the SXR derivative flux,
suggesting that the electron beam is accelerated during the
impulsive phase of the C2.8 flare. The context image is the radio
dynamic spectrum captured by EOVSA in the high-frequency range of
1$-$18~GHz, and it shows a radio burst accompanied by the first
pulsation, confirming the presence of accelerated electrons.

\begin{figure}
\centering \noindent\includegraphics[width=\linewidth]{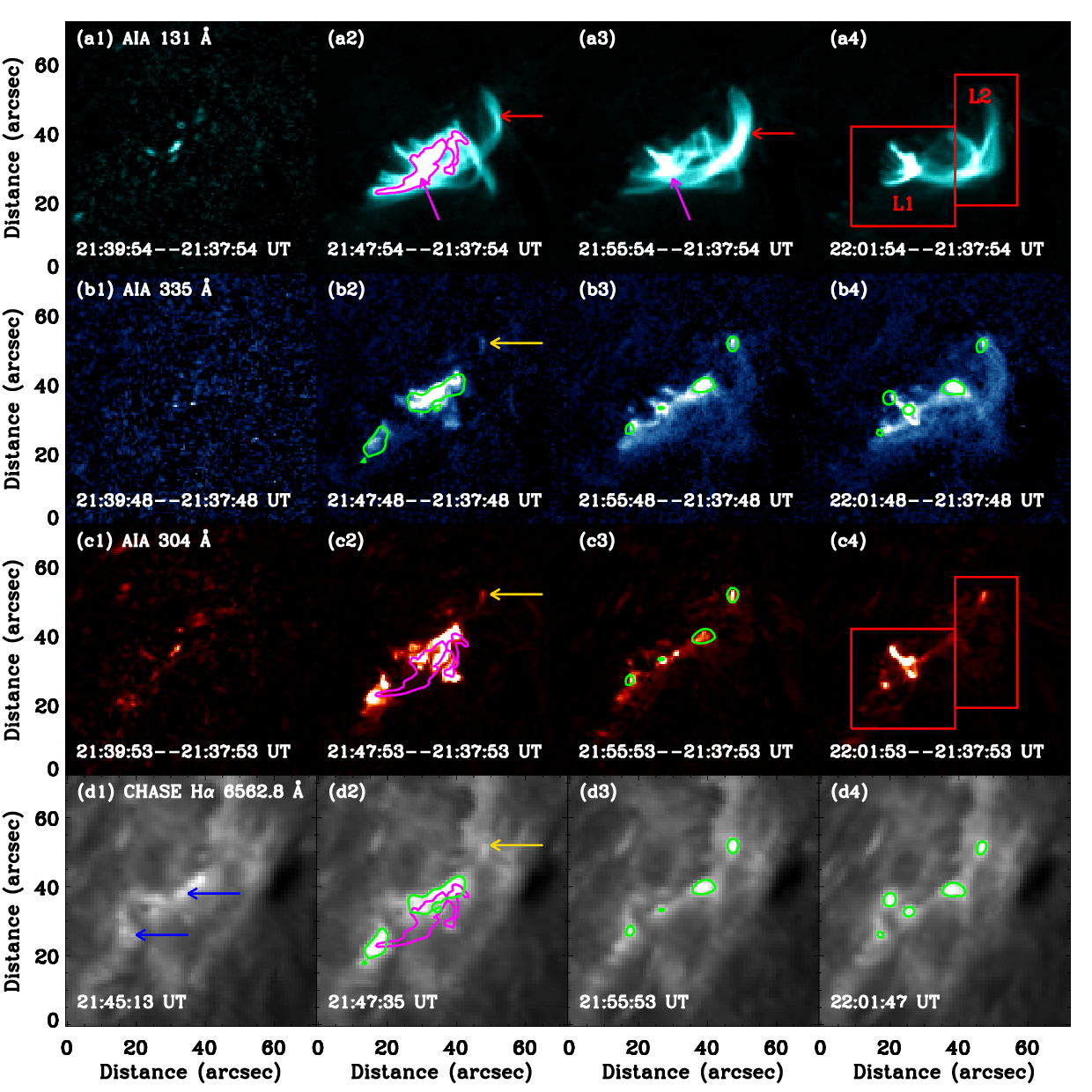}
\caption{Multiwavelength images with a FOV of
$\sim$72$^{\prime\prime}$$\times$72$^{\prime\prime}$ during the C2.8
flare. Panels~(a1)$-$(c4): Base different images measured by SDO/AIA at
131~{\AA}, 335~{\AA}, and 304~{\AA}. Panels~(d1)$-$(d4) H$\alpha$ images
observed by CHASE. The magenta contour represents the hot plasma
emission from AIA~131~{\AA}, while the green contours are the plasma
radiation from H$\alpha$. The magenta an red arrows mark the flare
loops, while the blue and gold arrows outline the flare kernels. The
red rectangles contain two hot loop systems (L1 and L2) at
AIA~131~{\AA}. \label{img}}
\end{figure}

Figure~\ref{img} gives a  close look at the source region of the
flare QPP, showing the multiwavelength images with a small FOV of
$\sim$72$^{\prime\prime}$$\times$72$^{\prime\prime}$ at four time
instances measured by SDO/AIA at 131~{\AA} (a1-a4), 335~{\AA}
(b1-b4), and 304~{\AA} (c1-c4), and CHASE H$\alpha$ at 6562.8~{\AA}
(d1-d4). Here, the different AIA-based images are given to show the
flare structure clearly, and the initial image is taken from the
image   captured two minutes before the flare onset. The H$\alpha$
raw images are given, since the CHASE begins to observe the Sun
after the flare onset and there are no pre-flare images.
Panels~(a1), (b1), and (c1) show the different AIA  images at the
flare onset time, and they all reveal very weak radiation in the
flare region. Panel~(d1) presents the first H$\alpha$ image measured
by CHASE during its observation for the C2.8 flare, and it exhibits
some H$\alpha$ kernels, as marked by the blue arrows. Panels~(a2),
(b2), (c2), and (d2) show the AIA and CHASE images during the first
pulsation (1). A group of hot loop-like structures can be clearly
seen in the wavelength of AIA~131~{\AA}, which is regarded as the
flare loop system, and is  outlined by the magenta contour (arrow).
Meanwhile, a weak loop system is also seen (marked by the red
arrow). On the other hand, two strong ribbon-like features that are
connected by the flare loop system (i.e., L1) appear in wavelengths
of AIA~335~{\AA} and 304~{\AA}, and the CHASE H$\alpha$ line center.
Moreover, a weak kernel-like structure (indicated by a gold arrow)
appears in these three images, which is connected by a weak hot loop
(i.e., L2). During the second pulsation (2) of the C2.8 flare, two
groups of loop-like structures are seen in the high-temperature EUV
channel, as indicated by the magenta contours in panel~(a3). These
middle- and low-temperature channels exhibits at least four
kernel-like structures, but they are a bit weaker, as indicated by
the green contours in panels~(b3), (c3), and (d3). Interestingly,
two hot loop systems connect these four kernels, which may be
regarded as flare footpoints. During the third pulsation (3) of the
C2.8 flare, the two hot loop systems and their corresponding ribbons
or footpoints can also be seen in EUV and H$\alpha$ images, as shown
in panels~(a4), (b4), (c4), and (d4). All these observations suggest
that the C2.8 flare shows multiple ribbons, at least four flare
ribbons. Then, two regions (red rectangles) that contain two flare
loop systems are separated to integrate their intensity curves.

\begin{figure}
\centering \noindent\includegraphics[width=\linewidth]{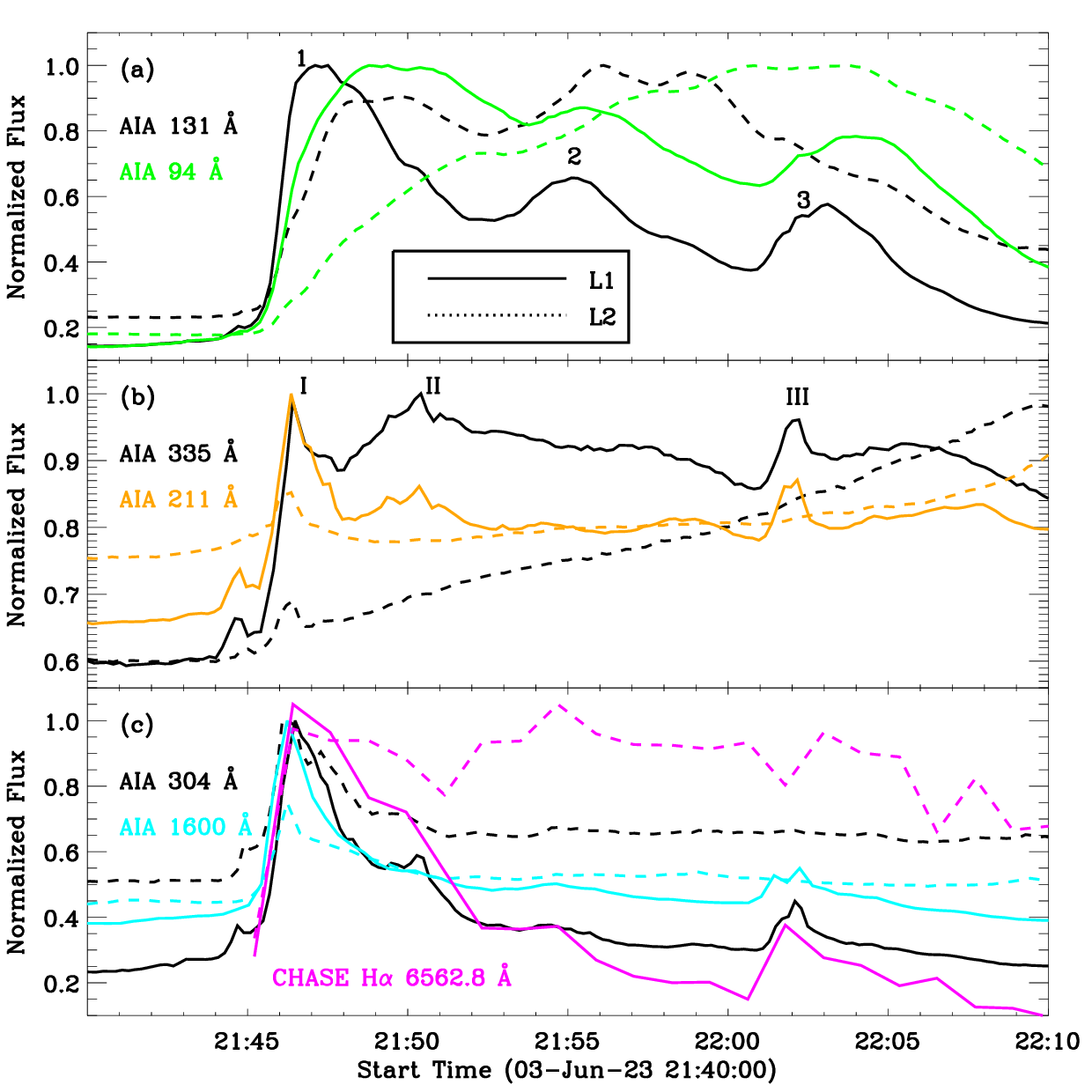}
\caption{Local light curves measured by SDO/AIA and CHASE~H$\alpha$.
The solid and dashed curves are integrated over the flare loop
regions of L1 and L2, respectively. \label{local}}
\end{figure}

Figure~\ref{local} shows the time series of the AIA intensity at two
loop-dominated regions. The solid line is integrated over from the
main strong loop system (L1), while the dashed line is integrated
over another weak loop system (L2). Panel~(a) presents the light
curves in high-temperature EUV channels, such as AIA~131~{\AA} and
94~{\AA}. Similarly to what has been seen in the SXR/XUV channels,
both the AIA~131~{\AA} and 94~{\AA} fluxes at the main strong loop
system (L1) show three pronounced pulsations during the C2.8 flare,
and they also appear to exhibit a decreasing trend, as indicated by
the solid black and green line profiles. On the other hand, the
AIA~131~{\AA} flux at the weak loop system (L2) also have three
small peaks, while the AIA~94~{\AA} flux at L2 loop mainly shows an
increasing trend. The similar monotonous growth can also be found at
the weak L2 loop system in wavelengths of AIA~335~{\AA} and
211~{\AA}, as shown by the dashed line curves in panel~(b). However,
such monotonic growth cannot be seen in the  AIA~304~{\AA} and
1600~{\AA} fluxes, nor can it be seen in the H$\alpha$~6562.8~{\AA}
light curve at the L2 loop system, as indicated by the dashed line
profiles in panel~(c). Figure~\ref{local}~(b) and (c) also
demonstrate that the light curves at the L1 loop system only have
one main pulsation (i.e., pulsation~1). Although the light curves at
AIA~335~{\AA} and 211~{\AA} appear as two subpeaks (I and II),  they
are still in the time interval of pulsation~1. We also note that
there is also a subpeak III during   pulsation~3 at the L1 loop in
almost all the observed wavelengths, but they are much weaker than
the first pulsation~1, which is different from that in the  AIA
high-temperature channels. All these observations suggest that the
three pulsations are basally from the flare area dominated by the
strong loop system.

\section{Discussions}
Based on the multi-instrument observations, we investigated the
flare QPP with a long period in wavelengths of SXR/XUV and
high-temperature EUV. The flare QPP was first observed in the
full-disk light curves in SXR/XUV channels of GOES~1$-$8~{\AA},
SXDU~1$-$10~{\AA}, ESP~1$-$70~{\AA,} and LYRA~1$-$200~{\AA}. There
are three successive and significant pulsations with a long
duration. The three pulsations in GOES~1$-$8~{\AA} flux are peaked
at about 21:47~UT, 21:55~UT, and 22:02~UT. The average period is
about 7.5~minutes. We also note that the peak time of each pulsation
occurs later  with increasing detected channels, which may because
the certain waveband emission might be dominated in a different time
of the flare, and it is consistent with our previous observations
\citep[e.g.,][]{Li23}. Then, the flare QPP with three successive
pulsations was also detected in the local light curves that are
integrated over the flare region at the high-temperature EUV
wavebands of AIA~131~{\AA} and 94~{\AA}, indicating that the QPP
feature   originates from the same flare region. However, the flare
QPP could not be observed in the GOES~1$-$8~{\AA} derivative flux,
which is regarded as the HXR flux according to the Neupert effect
\citep[e.g.,][]{Neupert68,Li24}, since we could not find the
available HXR light curve during the C2.8 flare. Moreover, the flare
QPP was also not detected in the radio emission,  at high or low
frequencies, and it was not seen in the local light curves at the
middle- and low-temperature AIA channels and the H$\alpha$ line
center. On the other hand, the light curves in these channels show
one pronounced peak that corresponds to   pulsation~1, suggesting
that   pulsation~1 may be caused by the electron beam accelerated by
the magnetic reconnection, as described in the standard 2D
reconnection model \citep{Priest02}. Instead, the  two later
pulsations might be due to the interaction of some blended loops
since we can find a series of hot loops in the AIA~131~{\AA} images,
which are regarded as flare loop systems, as seen in the online
animation.

The flare QPP with three successive pulsations in multiple
wavelengths suggests three energy-releasing processes at a long
period during the C2.8 flare. We note that the three pulsations in
the wavelength of AIA~94~{\AA} appear as an increasing trend; that is,
the peak intensity of pulsation~3 is obviously larger than that of
pulsation~1 (Figure~\ref{over}~a). This feature is definitively
different from that in the wavelength of AIA~131~{\AA},
GOES~1$-$8~{\AA}, and SXDU~1$-$10~{\AA}, all of which show a
decreasing trend. We then decomposed the whole flare region into two
loop-dominated regions (L1 and L2) by using the AIA images in EUV/UV
wavelengths. The strong loop-dominated region is considered   the
main flare region, and the light curves at AIA~131~{\AA} and
94~{\AA} that integrated over this main region show a very similar
decreasing trend, suggesting that the strong loop-dominated region
(L1) is the generated source of the flare QPP. While the weak
loop-dominated region (L2) is considered   the accompanying flare
area, and the light curves at AIA~94~{\AA}, 335~{\AA}, and 211~{\AA}
that integrated over this accompanying region reveal a trend of
monotonic growth. Therefore, the light curves at the middle-temperature
EUV channels show an increasing trend. It should be noted that the
middle-temperature channels are the EUV data in wavelengths of
AIA~94~{\AA}, 335~{\AA}, and 211~{\AA}, which is distinguished from
the high-temperature channel at AIA~131~{\AA} and the
low-temperature channels at AIA~171~{\AA}, 304~{\AA}, 1600~{\AA},
and CHASE~H$\alpha$. The light curves at the  low-temperature channels
only show one pronounced pulsation at the strong loop-dominated
region, similar to previous observations at the whole flare region
in the radio emission.

The flare QPP under study is different from previous flare QPPs
observed in SXR/EUV channels. Here, we reported a QPP event within a
long period of about 7.5~minutes during the entire flare
phases (i.e., from the impulsive phase through the main phase to the
decay phase). The previous QPPs events were often observed in
multiple channels of SXR, EUV, HXR, and radio/microwave during
solar--stellar flares, and their periods were commonly on a
timescale of seconds or minutes
\citep[e.g.,][]{Nakariakov18,Kolotkov21,Hong21,Lid22,Zimovets22,Mehta23,Corchado24}.
Moreover, the flare QPPs were frequently related to nonthermal
particles periodically accelerated by repetitive magnetic
reconnections, especially when they were detected
simultaneously in HXR and microwave fluxes during the impulsive
phase of solar flares \citep[e.g.,][]{Yuan19,Karampelas23,Li24b}. The
very long-period pulsations (VLPs) with periods of
$\sim$8$-$30~minutes have also been detected in wavelengths of SXR
\citep{Tan16} and H$\alpha$ \citep{Li20b}. However, those VLPs were
found before the onset of major solar flares rather than during the
solar flares themselves. Therefore, these VLPs might be some recurrences of
small flares before the major flare. In this case, the three
pulsations were located in the same region dominated by a group of
loop-like structures, which is regarded as one flare (see
also the solar monitor). The flare QPPs were also reported during the
entire phases of a solar flare, but they usually showed various
periods in different flare phases
\citep{Hayes19,Li20c,Li21,Collier23}. For instance, \cite{Hayes19}
reported the QPP with a period of $\sim$65~s in the impulsive phase
followed by the QPP with a period of $\sim$150~s in the decay phase,
which was also different from our case. In summary, we first reported a
long-period QPP during the entire flare phase in wavelengths of
SXR, XUV, and high-temperature EUV.

\begin{figure}
\centering \noindent\includegraphics[width=\linewidth]{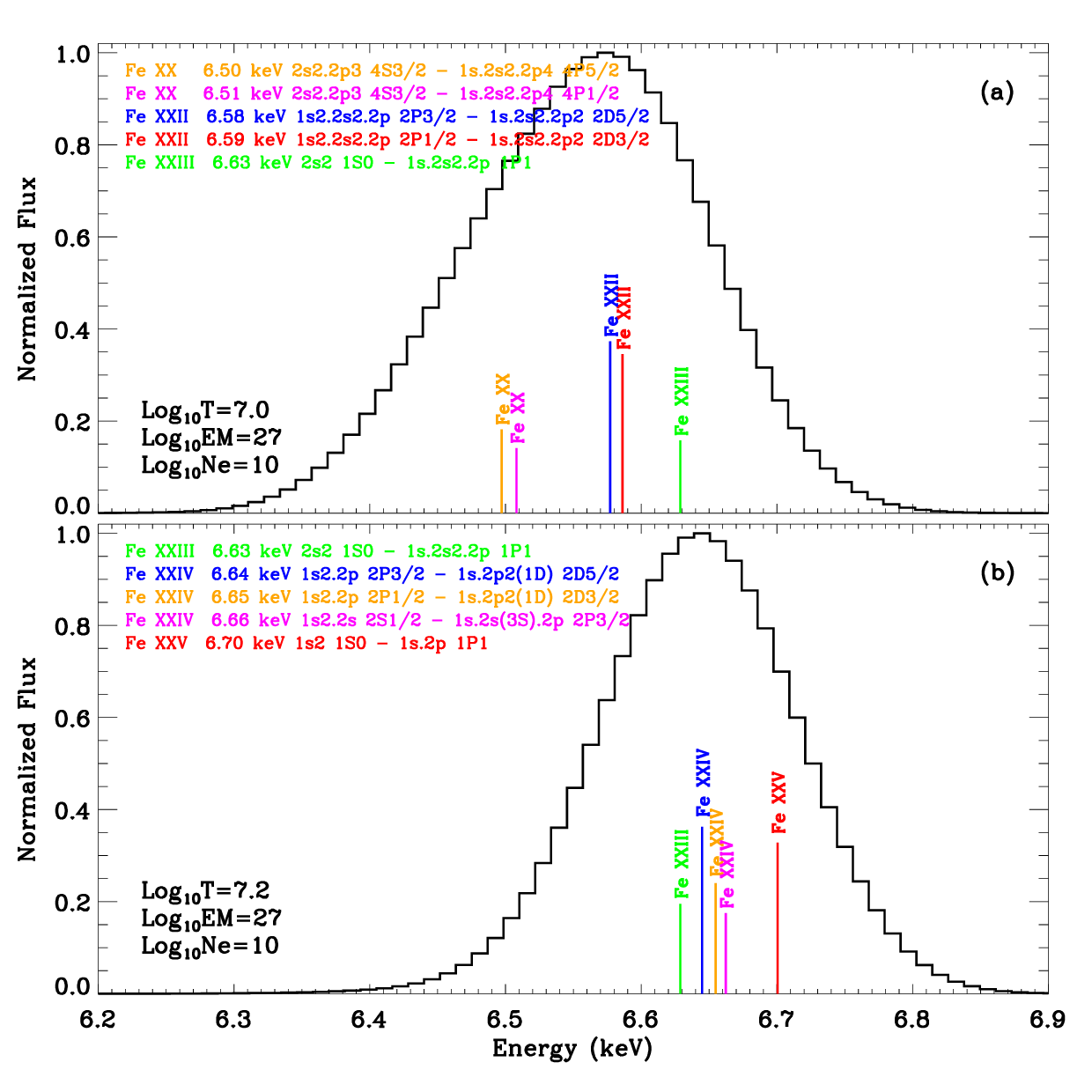}
\caption{Synthetic SXR spectra in the energy range between 6.2~keV
and 6.9~keV from the CHIANTI database at plasma temperatures
(Log$_{10}$T) of 7.0 (a) and 7.2 (b), respectively. The colored lines
mark the five strongest lines at the high-order ionized iron in the
given condition. \label{add}}
\end{figure}

The QPP behavior seen in the fitting center of Fe and Ca lines
mainly occurs because of the periodic variation of plasma
temperatures. Using a single Gaussian function with a linear trend,
we fitted the line profiles of Fe and Ca lines. Then, their fitting
peak intensities and line centers were extracted, both of which show
the QPP feature with a period of about 7.5~minutes. The fitting peak
intensity is the peak radiation of Fe and Ca lines during the C2.8
flare. However, it was impossible to determine the Doppler shift of
these two lines, although we   extracted the fitting line centers.
This is because   we could not identify their reference line
centers. Figure~\ref{add} presents the synthetic SXR spectra in the
energy range of 6.2$-$6.9~keV, which is computed from the CHIANTI
database \citep{DelZ21} for the flare condition at two temperatures:
Log$_{10}$T=7.0 (a) and Log$_{10}$T=7.2 (b). We note that the line
profile of Fe line is actually composed of a series of highly
ionized iron ions; that is, the dominant spectral lines are varied
from Fe XX to Fe XXV with increasing plasma temperature. So it is
impossible to determine the line center of each spectral line,
mainly due to the limitation of the spectral resolution of
MSS-1B/SXDU. In our case the spectral lines are simply called the Fe
line, and the periodic variation of fitting centers at the Fe line
is largely caused by the periodic disturbances of the plasma
temperature during the C2.8 flare. This was demonstrated by the
oscillation of the isothermal temperature of SXR-emitting plasmas
that were determined by two GOES SXR fluxes, as seen in
Figure~\ref{flux}~(a).

\section{Summary}
Using the spaced- and ground-based telescopes, MSS-1B/SXDU,
CHASE, GOES, PROBA2/LYRA, SDO/AIA, SDO/EVE, e-CALLISTO/NORWAY,
SWAVES, and EOVSA, we reported the observation of a QPP event during
a four-ribbon flare on 2023 June 03. Our main results are summarized
as follows:

(1) The flare QPP with an average period of about 7.5~minutes was
simultaneously detected in SXR, XUV, and high-temperature EUV
channels. It manifests as three pronounced pulsations in the light
curves at GOES~1$-$8~{\AA}, SXDU~1$-$10~{\AA}, ESP~1$-$70~{\AA},
LYRA~1$-$200~{\AA}, AIA~131~{\AA}, and 94~{\AA}; it can also be seen
in the time series of GOES temperature, the fitting peak, and the center
of the Fe and Ca lines.

(2) The flare QPP cannot be observed in the high- and
low-frequency radio emissions, and the middle- and low-temperature
EUV/UV wavelengths. Only one main pulsation (1) can be found in
the time series of GOES~1$-$8~{\AA} derivative, EOVSA~6.60~GHz,
NORWAY~1.23~GHz, and SWAVES~1.93~MHz, as well as the AIA~171~{\AA},
304~{\AA}, 335~{\AA}, and 1600~{\AA}, and H$\alpha$~6562.8~{\AA}.
This observational fact implies that the first pulsation is highly
associated with the nonthermal electron accelerated by magnetic
reconnection.

(3) AIA imaging observations show that the C2.8 flares have double
loop systems that connect four ribbons or kernels, and the QPP
feature is mainly from the flare area characterized by a strong loop
system.

(4) The QPP feature seen in time series of the fitting peak and
center of the  Fe and Ca lines may be due to the periodic variety
of the plasma temperature, causing the ion and calcium into
high-order ionizations.

(5) The flare QPP with three main pulsations suggest three
energy-releasing processes. The first energy-releasing process is
caused by the accelerated electron, while the later two
energy-releasing processes might be due to the loop-loop interaction
in the strong loop system.

\begin{acknowledgements}
We gratefully acknowledge the referee for his or her inspiring
comments.This work is funded by NSFC under grants 12250014, 12273101
12073081, the National Key R\&D Program of China 2021YFA1600502
(2021YFA1600500). This work was supported by the Macao Foundation.
D. Li is also supported by the Specialized Research Fund for State
Key Laboratories. This work is also supported by the Strategic
Priority Research Program of the Chinese Academy of Sciences, Grant
No. XDB0560000. We thank the teams of MSS-1B, CHASE, GOES, PROBA2,
SDO, e-CALLISTO, SWAVES, and EOVSA for their open data use policy.
The authors would also like to thank Drs. Chuan~Li and Ye~Qiu for
their discussions about the CHASE data, and Dr.~Sijie~Yu for
discussing the EOVSA data. The CHASE mission is supported by China
National Space Administration (CNSA).
\end{acknowledgements}



\begin{thebibliography}{}
\bibitem[Aschwanden(1987)]{Aschwanden87} Aschwanden, M.~J.\ 1987, \solphys, 111, 113.
\bibitem[Benz(2017)]{Benz17} Benz, A.~O.\ 2017, Living Reviews in Solar Physics, 14, 2.
\bibitem[Brosius \& Daw(2015)]{Brosius15} Brosius, J.~W. \& Daw, A.~N.\ 2015, \apj, 810, 45.
\bibitem[Chen et al.(2019)]{Chen19} Chen, X., Yan, Y., Tan, B., et al.\ 2019, \apj, 878, 78.
\bibitem[Chamberlin et al.(2009)]{Chamberlin09} Chamberlin, P.~C., Woods, T.~N., Eparvier, F.~G., et al.\ 2009, \procspie, 7438, 743802.
\bibitem[Collier et al.(2023)]{Collier23} Collier, H., Hayes, L.~A., Battaglia, A.~F., et al.\ 2023, \aap, 671, A79.
\bibitem[Corchado Albelo et al.(2024)]{Corchado24} Corchado Albelo, M.~F., Kazachenko, M.~D., \& Lynch, B.~J.\ 2024, \apj, 965, 16.
\bibitem[Didkovsky et al.(2012)]{Didkovsky12} Didkovsky, L., Judge, D., Wieman, S., et al.\ 2012, \solphys, 275, 179.
\bibitem[Del Zanna et al.(2021)]{DelZ21} Del Zanna, G., Dere, K.~P., Young, P.~R., et al.\ 2021, \apj, 909, 38.
\bibitem[Dominique et al.(2013)]{Dominique13} Dominique, M., Hochedez, J.-F., Schmutz, W., et al.\ 2013, \solphys, 286, 21.
\bibitem[Farhang et al.(2022)]{Farhang22} Farhang, N., Shahbazi, F., \& Safari, H.\ 2022, \apj, 936, 87.
\bibitem[Gary et al.(2011)]{Gary11} Gary, D.~E., Hurford, G.~J., Nita, G.~M., et al.\ 2011, AAS/Solar Physics Division Abstracts \#42
\bibitem[Hayes et al.(2019)]{Hayes19} Hayes, L.~A., Gallagher, P.~T., Dennis, B.~R., et al.\ 2019, \apj, 875, 33.
\bibitem[Hong et al.(2021)]{Hong21} Hong, Z., Li, D., Zhang, M., et al.\ 2021, \solphys, 296, 171.
\bibitem[Inglis et al.(2023)]{Inglis23} Inglis, A., Hayes, L., Guidoni, S., et al.\ 2023, \baas.
\bibitem[Janvier et al.(2015)]{Janvier15} Janvier, M., Aulanier, G., \& D{\'e}moulin, P.\ 2015, \solphys, 290, 3425.
\bibitem[Jiang et al.(2021)]{Jiang21} Jiang, C., Feng, X., Liu, R., et al.\ 2021, Nature Astronomy, 5, 1126.
\bibitem[Kaiser et al.(2008)]{Kaiser08} Kaiser, M.~L., Kucera, T.~A., Davila, J.~M., et al.\ 2008, \ssr, 136, 5.
\bibitem[Karlick{\'y} \& Ryb{\'a}k(2023)]{Karlicky23} Karlick{\'y}, M. \& Ryb{\'a}k, J.\ 2023, Universe, 9, 92.
\bibitem[Karampelas et al.(2023)]{Karampelas23} Karampelas, K., McLaughlin, J.~A., Botha, G.~J.~J., et al.\ 2023, \apj, 943, 131.
\bibitem[Kolotkov et al.(2021)]{Kolotkov21} Kolotkov, D.~Y., Nakariakov, V.~M., Holt, R., et al.\ 2021, \apjl, 923, L33.
\bibitem[Kupriyanova et al.(2020)]{Kupriyanova20} Kupriyanova, E., Kolotkov, D., Nakariakov, V., et al.\ 2020, Solar-Terrestrial Physics, 6, 3.
\bibitem[Lemen et al.(2012)]{Lemen12} Lemen, J.~R., Title, A.~M., Akin, D.~J., et al.\ 2012, \solphys, 275, 17.
\bibitem[Li et al.(2017)]{Li17} Li, T., Zhang, J., \& Hou, Y.\ 2017, \apj, 848, 32.
\bibitem[Li et al.(2022)]{Lic22} Li, C., Fang, C., Li, Z., et al.\ 2022, Science China Physics, Mechanics, and Astronomy, 65, 289602.
\bibitem[Li et al.(2020a)]{Li20a} Li, D., Kolotkov, D.~Y., Nakariakov, V.~M., et al.\ 2020a, \apj, 888, 53.
\bibitem[Li et al.(2020b)]{Li20b} Li, D., Feng, S., Su, W., et al.\ 2020b, \aap, 639, L5.
\bibitem[Li et al.(2020c)]{Li20c} Li, D., Lu, L., Ning, Z., et al.\ 2020c, \apj, 893, 7.
\bibitem[Li et al.(2021a)]{Li21a} Li, D., Warmuth, A., Lu, L., et al.\ 2021a, Research in Astronomy and Astrophysics, 21, 066.
\bibitem[Li et al.(2021b)]{Li21} Li, D., Ge, M., Dominique, M., et al.\ 2021b, \apj, 921, 179.
\bibitem[Li \& Chen(2022)]{Li22} Li, D. \& Chen, W.\ 2022, \apjl, 931, L28.
\bibitem[Li et al.(2022)]{Li22f} Li, D., Shi, F., Zhao, H., et al.\ 2022, Frontiers in Astronomy and Space Sciences, 9, 1032099.
\bibitem[Li(2022)]{Lid22} Li, D.\ 2022, Science in China E: Technological Sciences, 65, 139. doi:10.1007/s11431-020-1771-7
\bibitem[Li et al.(2023a)]{Li23} Li, D., Warmuth, A., Wang, J., et al.\ 2023a, Research in Astronomy and Astrophysics, 23, 095017.
\bibitem[Li et al.(2023b)]{Li23b} Li, D., Li, Z., Shi, F., et al.\ 2023b, \aap, 680, L15.
\bibitem[Li et al.(2024a)]{Li24} Li, D., Dong, H., Chen, W., et al.\ 2024a, \solphys, 299, 57.
\bibitem[Li et al.(2024b)]{Li24b} Li, D., Hong, Z., Hou, Z., et al.\ 2024b, \apj, 970, 77.
\bibitem[Masuda et al.(1994)]{Masuda94} Masuda, S., Kosugi, T., Hara, H., et al.\ 1994, \nat, 371, 495.
\bibitem[Masson et al.(2009)]{Masson09} Masson, S., Pariat, E., Aulanier, G., et al.\ 2009, \apj, 700, 559.
\bibitem[McLaughlin et al.(2018)]{McLaughlin18} McLaughlin, J.~A., Nakariakov, V.~M., Dominique, M., et al.\ 2018, \ssr, 214, 45.
\bibitem[McKevitt et al.(2024)]{McKevitt24} McKevitt, J., Jarolim, R., Matthews, S., et al.\ 2024, \apjl, 961, L29.
\bibitem[Mehta et al.(2023)]{Mehta23} Mehta, T., Broomhall, A.-M., \& Hayes, L.~A.\ 2023, \mnras, 523, 3689.
\bibitem[Milligan et al.(2020)]{Milligan20} Milligan, R.~O., Hudson, H.~S., Chamberlin, P.~C., et al.\ 2020, Space Weather, 18, e02331.
\bibitem[Motyk et al.(2023)]{Motyk23} Motyk, I.~D., Kashapova, L.~K., Setov, A.~G., et al.\ 2023, Geomagnetism and Aeronomy, 63, 1062.
\bibitem[Nakariakov et al.(2018)]{Nakariakov18} Nakariakov, V.~M., Anfinogentov, S., Storozhenko, A.~A., et al.\ 2018, \apj, 859, 154.
\bibitem[Nakariakov et al.(2019)]{Nakariakov19} Nakariakov, V.~M., Kolotkov, D.~Y., Kupriyanova, E.~G., et al.\ 2019, Plasma Physics and Controlled Fusion, 61, 014024.
\bibitem[Nakariakov \& Kolotkov(2020)]{Nakariakov20} Nakariakov, V.~M. \& Kolotkov, D.~Y.\ 2020, \araa, 58, 441.
\bibitem[Neupert(1968)]{Neupert68} Neupert, W.~M.\ 1968, \apjl, 153, L59.
\bibitem[Ning et al.(2022)]{Ning22} Ning, Z., Wang, Y., Hong, Z., et al.\ 2022, \solphys, 297, 2.
\bibitem[Parks \& Winckler(1969)]{Parks69} Parks, G.~K. \& Winckler, J.~R.\ 1969, \apjl, 155, L117.
\bibitem[Phillips(2004)]{Phillips04} Phillips, K.~J.~H.\ 2004, \apj, 605, 921.
\bibitem[Priest \& Forbes(2002)]{Priest02} Priest, E.~R. \& Forbes, T.~G.\ 2002, \aapr, 10, 313.
\bibitem[Samanta et al.(2021)]{Samanta21} Samanta, T., Tian, H., Chen, B., et al.\ 2021, The Innovation, 2, 100083.
\bibitem[Shen et al.(2022)]{Shen22} Shen, Y., Yao, S., Tang, Z., et al.\ 2022, \aap, 665, A51.
\bibitem[Shen et al.(2023)]{Shen23} Shen, J., Li, J., Huang, Y., et al.\ 2023, \apj, 950, 71.
\bibitem[Shi et al.(2023)]{Shi23} Shi, Y., Li, L., Chen, J., et al.\ 2023, Earth and Planetary Physics, 7, 125.
\bibitem[Tan \& Tan(2012)]{Tan12} Tan, B. \& Tan, C.\ 2012, \apj, 749, 28.
\bibitem[Tan et al.(2016)]{Tan16} Tan, B., Yu, Z., Huang, J., et al.\ 2016, \apj, 833, 206.
\bibitem[Tian et al.(2008)]{Tian08} Tian, H., Xia, L.-D., \& Li, S.\ 2008, \aap, 489, 741.
\bibitem[Tian(2017)]{Tian17} Tian, H.\ 2017, Research in Astronomy and Astrophysics, 17, 110.
\bibitem[Wang et al.(2014)]{Wang14} Wang, H., Liu, C., Deng, N., et al.\ 2014, \apjl, 781, L23.
\bibitem[Warmuth \& Mann(2016)]{Warmuth16} Warmuth, A. \& Mann, G.\ 2016, \aap, 588, A115.
\bibitem[Yan et al.(2018)]{Yan18} Yan, X.~L., Yang, L.~H., Xue, Z.~K., et al.\ 2018, \apjl, 853, L18.
\bibitem[Yan et al.(2022)]{Yan22} Yan, X., Xue, Z., Jiang, C., et al.\ 2022, Nature Communications, 13, 640.
\bibitem[Yuan et al.(2019)]{Yuan19} Yuan, D., Feng, S., Li, D., et al.\ 2019, \apjl, 886, L25.
\bibitem[Zhang(2024)]{Zhang24} Zhang, Q.\ 2024, Reviews of Modern Plasma Physics, 8, 7.
\bibitem[Zhao et al.(2023)]{Zhao23} Zhao, H.-S., Li, D., Xiong, S.-L., et al.\ 2023, Science China Physics, Mechanics, and Astronomy, 66, 259611.
\bibitem[Zhou et al.(2024)]{Zhou24} Zhou, X., Shen, Y., Yan, Y., et al.\ 2024, \apj, 968, 85.
\bibitem[Zimovets et al.(2021a)]{Zimovets21a} Zimovets, I., Sharykin, I., \& Myshyakov, I.\ 2021a, \solphys, 296, 188.
\bibitem[Zimovets et al.(2021b)]{Zimovets21b} Zimovets, I.~V., McLaughlin, J.~A., Srivastava, A.~K., et al.\ 2021b, \ssr, 217, 66.
\bibitem[Zimovets et al.(2022)]{Zimovets22} Zimovets, I.~V., Nechaeva, A.~B., Sharykin, I.~N., et al.\ 2022, Geomagnetism and Aeronomy, 62, 356.
\end{thebibliography}
\end{document}